\documentclass[12pt]{article}
\usepackage{amsmath}
\usepackage{amssymb}
\usepackage{graphicx,bbm,mathrsfs}
\usepackage{color}

\newcommand{\lsim}{\raisebox{-0.13cm}{~\shortstack{$<$ \\[-0.07cm] $\sim$}}~} 
\newcommand{\gsim}{\raisebox{-0.13cm}{~\shortstack{$>$ \\[-0.07cm] $\sim$}}~} 
\newcommand{\beq}{\begin{eqnarray}} 
\newcommand{\eeq}{\end{eqnarray}}

\begin{document}

\begin{flushright}LPT-Orsay--15--98 \end{flushright} 

\begin{center}

{\large\bf 
The LHC diphoton resonance and dark matter}

\vspace*{5mm}

{\sc Yann Mambrini, Giorgio Arcadi} and {\sc Abdelhak~Djouadi} 

\vspace{3mm}

{\small 
Laboratoire de Physique Th\'eorique,  CNRS and Universit\'e Paris-Saclay \\  
B\^at. 210, F--91405 Orsay Cedex, France \\
}
\end{center}

\begin{abstract}
A Higgs--like resonance with a mass of approximately 750 GeV has recently been ``observed" at the LHC in its diphoton decay. If this state is not simply a statistical fluctuation which will disappear with more data, it will have important implications not only for particle physics but also for cosmology. In this note, we analyze the implications of such a resonance for the dark matter (DM). Assuming a spin--$\frac12$  DM particle, we first verify that indeed the correct relic density can be obtained for a wide range of the particle mass and weak scale coupling, that are compatible with present data.  We then show that  the  combination of near future direct and indirect detection  experiments will allow to probe the CP--nature of the mediator resonance, i.e. check whether it is a scalar or a pseudoscalar like particle.  
\end{abstract}

\subsection*{1. Introduction}

In the searches performed at the new LHC run with a center of mass energy of 13 TeV and 
a few fb$^{-1}$ accumulated data, the ATLAS and CMS collaborations have reported
the observation of a diphoton resonance at an invariant mass of about $750\;$GeV
\cite{annonce}. Of course,  the significance of the signal is rather modest, only slightly above three standard deviations in the case of ATLAS and less in the case of CMS, and the effect could thus simply be a statistical fluctuation which will go away, like many past excesses, with more accumulated data. It is nevertheless tempting to consider the possibility that it is due to new physics. If true, this will have far reaching consequences not only for particle physics but also for astrophysics and cosmology as it will be discussed in this note.  

But before that, let us briefly summarize the information that is available so far on the possibly new resonance. First, since the particle has been observed in its diphoton decay mode, it is very likely that it has zero spin, since a spin--1 state decaying into two photons is ruled out by the Landau--Yang theorem \cite{LandauYang}. Second, the large production cross section of ${\cal O}(10~{\rm fb})$ indicates that it has presumably been produced through its couplings to gluons. Indeed, since the luminosity for quarks increases less steeply with the collider energy than the one for gluons \cite{collider-reach}, the resonance should have been observed at the previous LHC run with lower energies if it were produced in quark--antiquark annihilation with such rates.  A third information is that its total decay width of about 50 GeV is so large, that the resonance must couple strongly to the particles in which it decays. These daughter particles are unlikely $W,Z$ bosons, leptons and even light quark jets on which present  exclusion bounds  are severe \cite{PDG,Latest-13TeV}. They should be final states that are too difficult to search for such as, for instance, top quarks that are subject to a huge QCD background or states that decay partly or fully into invisible. Hence, to summarize, the spin--zero resonance should have loop induced but  strong couplings to gluons and photons and should decay primarily into final states that are difficult to observe.  

Such a resonance has thus the ideal properties to play a prominent role in the physics of the
particles that form the dark matter (DM) in the universe \cite{DM-review} and which are the most wanted    particles in both accelerator based experiments and astrophysical experiments. Indeed, the present wisdom summarised by the weakly interacting massive particle or WIMP paradigm, is that an electrically neutral particle with a mass in the few 10 GeV to few hundred GeV range and interacting weakly with the visible sector, should be stable at cosmological scales and accounts for the DM with  a relic abundance that has been precisely measured by the WMAP and PLANCK satellites \cite{WMAP,PLANCK}.  

In this brief note, we investigate the possibility that the observed diphoton resonance mediates the interactions of a spin--$\frac12$ DM particle. We will work in a rather model independent framework in which the new particle content associated to both the resonance and the DM states is not specified and the interactions are described by effective operators. We first show that the measured value of the cosmological relic density can be reproduced for a wide range of the DM particle masses and couplings. We then discuss the present bounds and the future sensitivities that can be achieved on the these parameters from astrophysical detection experiments, both  direct such as XENON \cite{XENON} and LUX \cite{Akerib:2013tjd} and more precisely in perspective of the new LZ project \cite{LZ}. We also study  indirect searches at the HESS \cite{Hess} and FERMI \cite{Ackermann:2013yva} experiments. The complementarity of the approaches is demonstrated as they are differently sensitive to the CP nature of the mediator resonance.

%%%%%%%%%%%%%%%%%%%%%%%%%%%%%%%%%%%%%%%%%%%%%%%%%%%%%%%%%%%%%%%%%%%%%%%%%%%%%%%%%%%%%%%
\subsection*{2. Effective interactions of the diphoton resonance }

We start by discussing the interactions of the diphoton resonance with the SM and DM particles. For simplicity, we consider a Majorana DM particle in our work, but the generalization to a Dirac fermion is straightforward. The interactions will be described in a model independent way in terms of effective operators for given $J^{\rm P}$ spin--parity     quantum numbers of the $\phi$ resonance. Two widely different possibilities need to be considered. 

A first one is that the $\phi$ particle has no direct couplings  to SM fermions. In this case, its interactions with gluons and electroweak gauge bosons are given by the following two Lagrangians. In the case of a CP--even $0^+$ particle, one has~\cite{Barbieri:2010nc}: 
\beq
{\cal L}_{0^+}= \frac{c_1}{\Lambda} \phi F_{\mu \nu} F^{\mu \nu} + 
\frac{c_2}{\Lambda} \phi W^{\mu \nu} W_{\mu \nu} + \frac{c_3}{\Lambda} 
\phi G^a_{\mu \nu}G_a^{\mu \nu} +g_\phi \phi \bar \chi \chi + m_\psi \bar \chi \chi .
\label{L0S}
\eeq
with $F_{\mu \nu}= (\partial_\mu Y_\nu - \partial_\nu Y_\mu)$ the field strength of 
the $Y_\mu$ hypercharge SM gauge field; the same holds for the SU(2) $W_\mu$ fields 
and the SU(3) $G_\mu$ fields. In the case where the mediator of the interaction $\phi$ is 
a CP--odd or pseudoscalar $0^-$  particle, one would have instead  \cite{Barbieri:2010nc}
\beq
{\cal L}_{0^-}= \frac{c_1}{\Lambda} \phi F_{\mu \nu} \tilde F^{\mu \nu} + 
\frac{c_2}{\Lambda} \phi W^{\mu \nu} \tilde W_{\mu \nu} + \frac{c_3}{\Lambda} \phi G^a_{\mu \nu} \tilde G_a^{\mu \nu} +i g_\phi \phi \bar \chi \gamma^5 \chi + m_\psi \bar \chi \chi .
\label{L0P}
\eeq
with $\tilde F_{\mu \nu} = \epsilon^{\mu \nu \rho \sigma} F_{\rho \sigma}$ and likewise
for the SU(2) and SU(3) gauge fields. On should note that while for LHC physics the CP nature of the $\phi$ resonance should not matter much, it is very important when it comes to dark matter searches.

The effective couplings of the $\phi$ state to the SM gauge bosons can be then written as
\beq
c_{\gamma \gamma}= c_1 \cos^2 \theta_W + c_2 \sin^2 \theta_W \, ,  \ 
c_{ZZ}=c_1 \sin^2 \theta_W + c_2 \cos^2 \theta_W \, , c_{WW}=c_2 , \,  c_{gg}=c_3
\eeq

There is also the possibility that the mediator $\phi$ has direct couplings to SM 
fermions.  As a bilinear term of the form $\phi \bar f f$ is not gauge invariant and explicitly breaks the SM gauge symmetry, we will assume an effective coupling of the $\phi$ particle to fermions given by the effective Lagrangians in the scalar and pseudoscalar cases 
\beq
{\cal L}_1 = {\cal L}_{0^+} + c_f \frac{m_f}{\Lambda} \phi \bar f f  \ \ {\rm or} \ \ 
{\cal L}_1 = {\cal L}_{0^-} + i c_f \frac{m_f}{\Lambda} \phi \bar f \gamma_5 f \ .
\label{L1M}
\eeq
where the Yukawa coupling is proportional to the fermion mass. This is typically 
what occurs, in particular in the Standard Model (SM) and its two Higgs doublet model 
(2HDM) extensions in which one has $\Lambda \approx v$, where $v = 246$ GeV is the vacuum expectation value of the SM Higgs field. The top quark should be then the particle that largely dominates the coupling to the scalar and pseudoscalar mediators. These two situations can be reproduced in the effective Lagrangians of eqs.~(\ref{L0S},\ref{L0P},\ref{L1M}) above by simply setting  $\Lambda=v$. For another microscopic ultraviolet version of such an extension,  see for instance Ref.~\cite{Mambrini:2015nza}. 

As the mediator resonance is presumably produced at the LHC in the gluon--gluon fusion process, $gg\to \phi$, and decays into two photons, $\phi \to \gamma \gamma$, one should have strong couplings to both particles in order to attain the cross section of ${\cal O}(10~{\rm
fb})$ that has been measured by ATLAS for instance. From the value of the cross section times the branching ratio, one in principle has a handle on the product $c_{\gamma \gamma} \times 
c_{gg}$. 

There is nevertheless a  third ingredient that enters the game since the $\phi \to \gamma \gamma$ branching ratio depends also on the $\phi$ total decay width. If the $\phi$ couplings to top quark are significant, the partial width $\phi \to t \bar t$ can completely saturate the total decay width, $\Gamma(\phi \to t \bar t) \approx \Gamma_\phi$ and
the branching fractions for the $\phi \to \gamma\gamma $ and even $\phi \to gg$ decays are  small. In turn, if there are no direct couplings to the top quark, these two decays and in particular the one involving strong interaction, will be the most important ones. This
is particularly true as the partial width grows like $M_\phi^3$ in this case. Thus, one 
can have scenarii in which $\Gamma(\phi \rightarrow gg) = 2 c_3^2 M_\phi^3/ (\pi \Lambda^2)
\approx 40$ GeV which sets a limit on $c_3/\Lambda$. 

Furthermore, there is another model dependence that is introduced by the possibility 
that the DM particle $\chi$ is relatively light, $m_\chi < \frac12 M_\phi$, allowing 
for the invisible decay channel  $\phi \to \chi \chi$ to occur with a partial width given by $\Gamma(\phi \rightarrow \chi \chi) = 2 g_\phi^2 M_\phi (1 - 4 m_\chi^2/ M_\phi^2)^{3/2}/(8 \pi)$.  Depending on the magnitude of the Yukawa couplings, one obtains either small
or large invisible decay widths.  For instance, if only the $gg$ and invisible decay channels are relevant (and hence one ignores the fermions Yukawa couplings), one obtains  for the invisible decay rate, ${\rm BR} (\phi \to {\rm inv})= [1+ 8 c_3^2 M_\Phi^2/(g_\phi^2 \Lambda^2)]^{-1}$.

\begin{figure}[!h]
\begin{center}
\mbox{\hspace*{-1mm} 
 \includegraphics[width=0.42\linewidth]{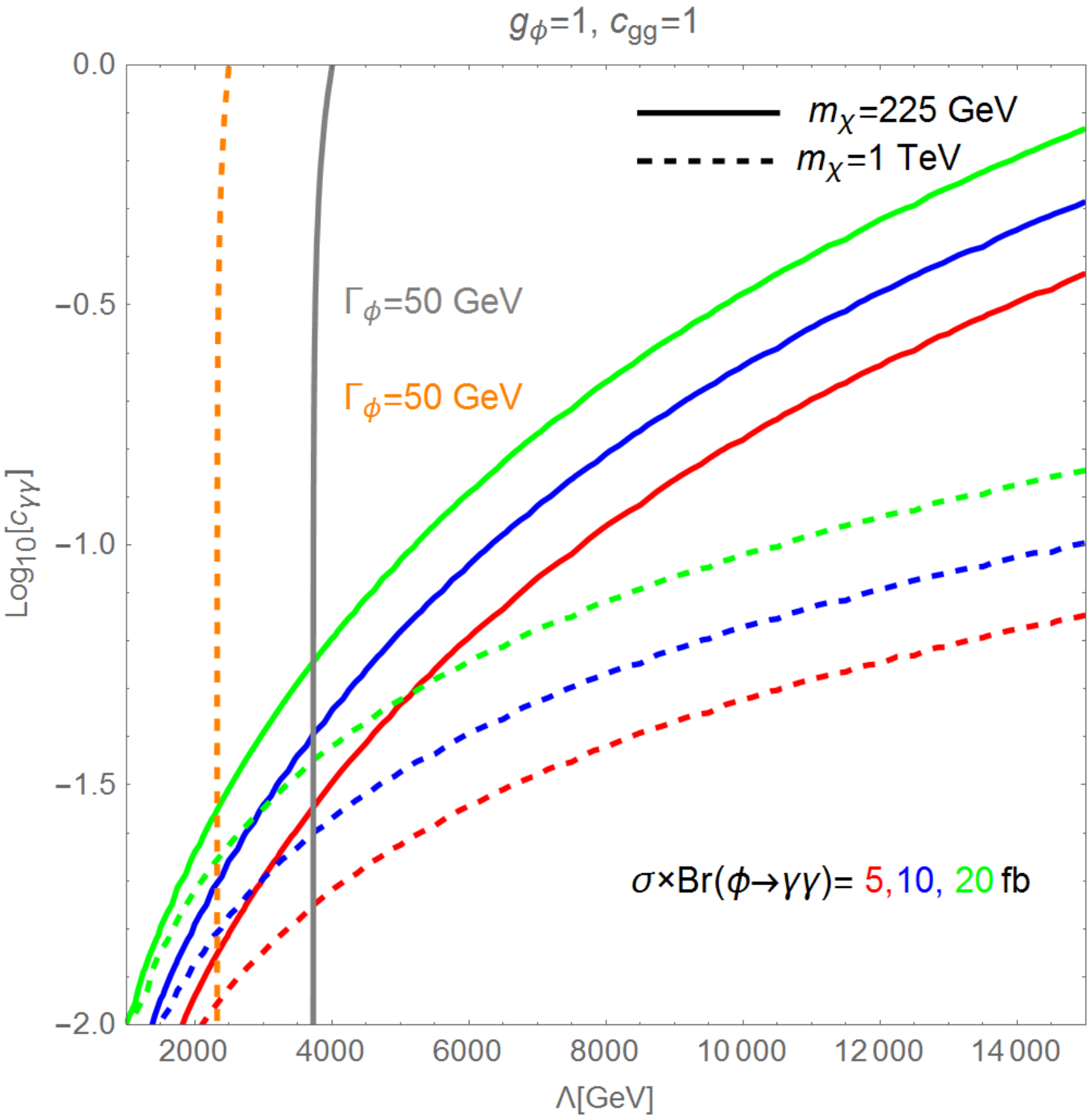}~~~~
 \includegraphics[width=0.42\linewidth]{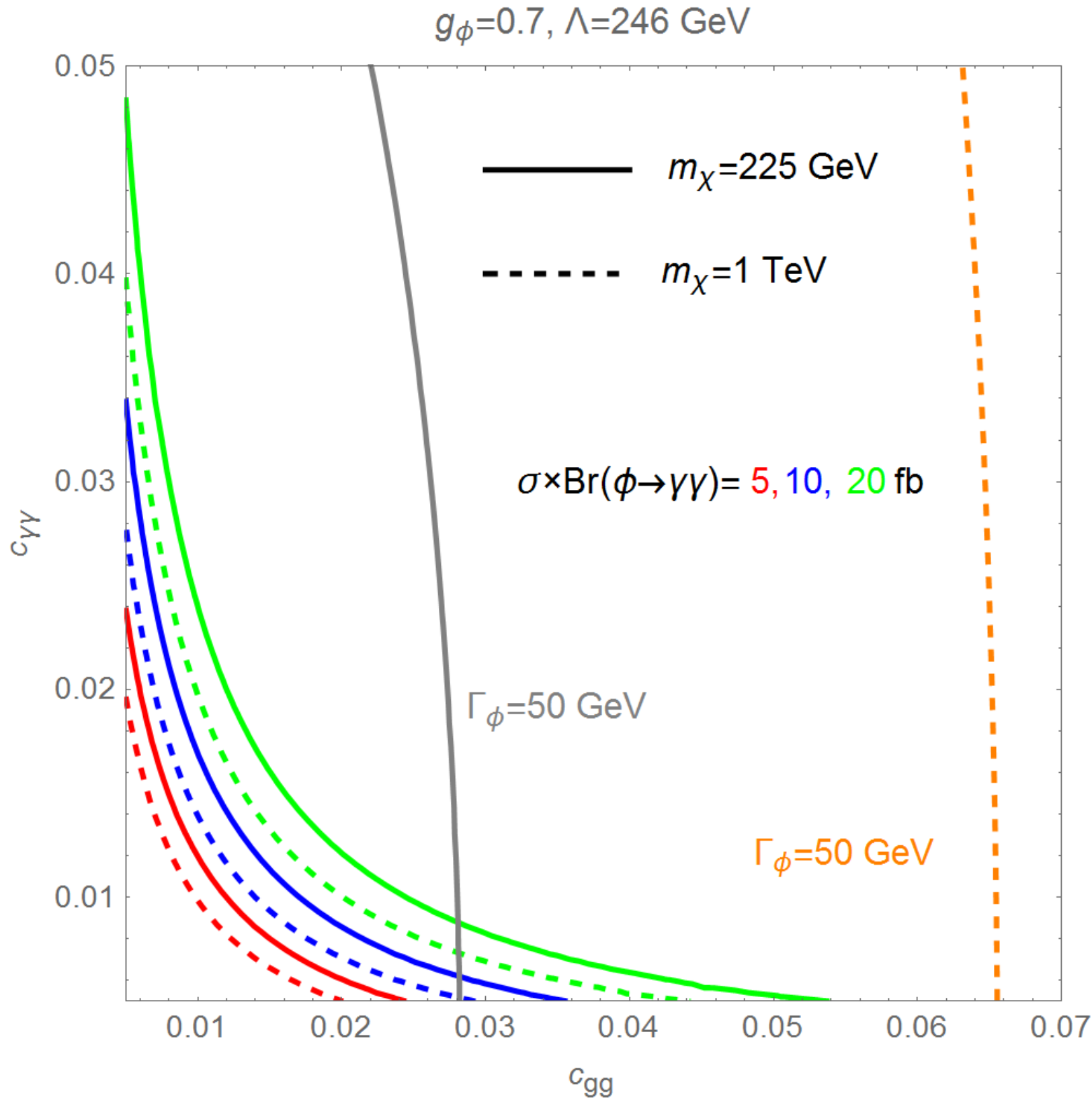}
}
\vspace*{-3mm}
\caption{\small Left panel: the result of the fit of the LHC data on the $\phi$ resonance mass, its cross section times branching ratio into two photons and its total width, in
the plane $[\Lambda,  \log_{\rm 10} c_{\gamma \gamma}]$ for a gluon  effective parameter  $c_{gg}=1$. We set $g_\phi=1$ and assume  $m_\chi=225$ GeV and $m_\chi=1$ TeV. Right panel: 
the same as before in the plane $[c_{gg},  c_{\gamma \gamma}]$ for $g_\phi=0.7$ and $\Lambda=246\,\mbox{GeV}$. } 
\end{center}
\label{Fig:LHC-fit}
\vspace*{-.9cm}
\end{figure}

Using all this information, one can fit some of or all the parameters of the effective 
Lagrangians that  describe the $\phi$ interactions.  The results of such a fit are
shown in Fig.~1. In the left panel, the results are given in the bi-dimensional plane 
formed by the scale $\Lambda$ and $\log_{\rm 10} c_{\gamma \gamma}$, which describes the photon couplings of the $\phi$ resonance, with the coupling $c_{gg}$ responsible for the production of the $\phi$ resonance being set to one. In the right panel, the plane 
$[c_{gg},c_{\gamma \gamma}]$ is instead considered with $\Lambda=246$ GeV. In both cases we have assumed the presence of a Yukawa coupling $c_t=1$  with the top quarks and a coupling $g_\phi$ of order unity. Both panels report the three values $\sigma(gg \to \phi) \times {\rm BR}(\phi \to \gamma \gamma)=5,10$ and 20 fb for the diphoton production rate, together with an isocontour for a total width of $\Gamma_\phi =50$ GeV. Two values have been chosen for the  mass of the DM particle: $m_\chi=225$ GeV which  corresponds to a sizable contribution of the decay $\phi \rightarrow \chi \chi$ to the total width and $m_\chi=1$ TeV in which there are only visible decay channels of $\phi$ instead.   

It can be argued that for high values of $\Lambda$, relatively large values of the $c_i$ parameters are needed to obtain values of $\sigma(gg \to \phi) \times {\rm BR}(\phi \to \gamma \gamma)$ of the order of 10 fb; on the contrary, values of $c_i$ of the order of $10^{-2}$ are sufficient if $\Lambda$ is set to the EW scale. Among the possible set of parameters that reproduce the experimental data,  we will focus in our study on the following two benchmarks with $M_\phi=750$ GeV: 
\begin{align} 
\label{Eq:parameters}
& \Lambda = 3 ~ \rm{TeV}, \, ~~~ c_{WW}=0, \, c_t=0, \,  c_{\gamma \gamma} \approx 0.2,~ \, c_{gg} \approx 1  \\
& \Lambda=246~\rm{GeV}, \, c_{WW}=0, \, c_t=1 ,\, c_{\gamma \gamma} \approx 0.01, \, c_{gg} \approx 0.01 
\label{Eq:parameters2}
\end{align}
The first set of parameters eq.~(\ref{Eq:parameters}) assumes that a significant effective point--like coupling of the resonance is obtained from new physics at a scale $\Lambda$ that is far above the electroweak scale with $v\approx 246$ GeV. The second set
eq.~(\ref{Eq:parameters2}) corresponds, instead, to the parameters that can be obtained in a realistic and ultraviolet two Higgs doublet model in which the diphoton resonances are in fact the CP--even and CP--odd additional $H$ and $A$ bosons. For values of order unity of the parameter $\tan\beta$, the ratios of the vev of the two Higgs fields, the Yukawa couplings of these states to the top quarks are large, $m_t/v = {\cal O}(1)$. The $\phi=H,A$ resonances will mainly decay into top quark pairs, giving a total width which is of the order of $\Gamma_\phi=40$ GeV for a mass around $M_\phi=750$ GeV. The $\phi$ couplings to gluons and photons are then generated via top quarks loops but additional matter content is necessary in order to enhance the $\gamma\gamma$ signal. We also note that in the so-called alignment limit
of this 2HDM, even if $\phi$ is a scalar state, it does not couples to $WW$ and $ZZ$ bosons and we have set $c_{WW}=0$ in order to reproduce this feature. Such a scenario will be investigated in detail in a forthcoming publication \cite{DM-Orsay}. 

With these parameters we are now ready to study the implications for DM starting with the cosmological relic density and, after that, the direct and indirect detection rates.

%%%%%%%%%%%%%%%%%%%%%%%%%%%%%%%%%%%%%%%%%%%%%%%%%%%%%%%%%%%%%%%%%%%%%%%%%%%%%%%%

\subsection*{3. Implications for the DM abundance and detection rates}

%\subsubsection*{3.1 Relic density}

In recent years, the WMAP \cite{WMAP} and PLANCK \cite{PLANCK} satellites have set severe
constraints on the relic abundance of DM in the universe. If eventually combined with accelerator, direct and indirect detection constraints, the relic density can impose strong constraints on the coupling of the DM particle to the state that mediates its annihilation. This is for instance the case, when the mediators are the 125 GeV Higgs \cite{Hportal} or the  Z boson~\cite{Zportal}. However, in the case of the much heavier diphoton 750 GeV resonance, the relic abundance as precisely measured, sets in general the most important constraints. 

Assuming that the DM particles annihilate into SM particles through the $s$--channel exchange of the $\phi$ state, $\chi \chi \to XX$ where $X$ stand for gluons, weak bosons, photons as well as fermions in models with direct $\phi f\bar f$ couplings, we have  entered the complete set of Feynman rules describing these processes in the latest released version
of the program micrOmegas \cite{micromegas} which is the basic tool that calculates
the relic abundance.

We have calculated the relic density by scanning over the parameter space that is obtained when varying the couplings and the mass of the DM particle over a wide range. We nevertheless limited our DM mass to a maximum value of 3 TeV above which the effective theory approach is not valid anymore. Note that for masses above $m_\chi \gsim 750$ GeV, we have also included 
t--channel annihilation into two $\phi$ particles, $\chi \chi \to \phi \phi$, 
whose contribution is 10 to 20 \% of the total annihilation rate in the first scenario
for instance.

In the scenario in which $\phi$ has no couplings to SM fermions, only the gauge boson final states are present. One obtains for the annihilation cross sections of a scalar and pseudoscalar resonance into two photons and two gluons, the following expressions
\beq
\langle \sigma v \rangle_{\gamma \gamma}^{0^+} \simeq 
\frac{ 4 g_\phi^2 c_{\gamma \gamma}^2 m_\chi^4 v^2}{\pi \Lambda^2 (s-M_\phi^2)^2}\, , ~~ 
\langle \sigma v \rangle_{\gamma \gamma}^{0^-}= 
\frac{4 g_\phi^2 c_{\gamma \gamma}^2 s^2}{\pi \Lambda^2(s-M_\phi^2)^2} \, , ~~
\langle \sigma v \rangle_{gg} = 8 \left( \frac{c_{gg}}{c_{\gamma \gamma}}\right)^2
\langle \sigma v \rangle_{\gamma \gamma}
\label{Eq:sigv}
\eeq 
where we have omitted the total decay widths in the propagators; $s$ is the center of mass energy given by $s \simeq 4 m_\chi^2 + m_\chi^2 v^2$ with $v$ the velocity.  The dominant channel is obviously the gluonic final state. As a result of the large mass of the mediator $\phi$, the cross sections are strongly suppressed and are significant only for relatively large values of the coupling $g_\phi$. Moreover, in the case of the CP--even state exchange, the annihilation cross section is velocity suppressed, imposing even larger values to $g_\phi$. 

We display in Fig.~2  the parameter space allowed by the measurement of the relic density  \cite{PLANCK} in the scalar or CP--even (blue line) and in the pseudoscalar or CP--odd (red line) cases for $M_\phi \approx 750$ GeV and the values of the $c_i$ parameters given in eqs.~(\ref{Eq:parameters},\ref{Eq:parameters2}). One clearly  distinguishes three different regions depending on the $\chi$ mass range. 

In the large $\chi$ mass region, $m_\chi \gg M_\phi$, the annihilation cross section into gluons can be approximated by $\langle \sigma v \rangle_{gg}^S \simeq 2 g_\phi^2 c_{gg}^2 v^2/ (\pi \Lambda^2)$ in the scalar case. If one takes $v \simeq 0.3$ at decoupling time and $\Lambda \simeq 3$ TeV, one obtains 
\beq
\langle \sigma v \rangle_{gg}^{0^+} \simeq 2 g_\phi^2 c_{gg}^2 v^2/ (\pi \Lambda^2) 
\simeq \left(\frac{g_\phi}{0.2} \right)^2 c_{gg}^2~~ 10^{-26}{\rm cm^3 s^{-1}} \ , 
~~~ m_\chi \gg M_\phi. 
\eeq
which needs $g_\phi \simeq 0.4$--0.7 to fit the observed relic abundance in agreement with the numerical analysis that led to Fig.~2. For a pseudoscalar mediator, 
eq.~(\ref{Eq:sigv}) leads 
\beq
\langle \sigma v \rangle_{gg}^{0^-} \simeq \frac{32 g_\phi^2 c^2_{gg}}{\pi \Lambda^2} \simeq
 \left(\frac{g_\phi}{0.02} \right)^2 c_{gg}^2~~ 10^{-26} {\rm cm^3 s^{-1}} \ , 
~~~ m_\chi \gg M_\phi. 
\eeq
The velocity suppression factor $v^2$ being absent, a much lower value of $g_\phi$ is needed to fulfil the relic abundance constraint in the pseudoscalar case. Our analytical approximation  gives $g_\phi \simeq 10^{-2}-10^{-1}$, in  accord with the results obtained numerically using micrOmegas as one can see from Fig.~2. One notices that in the regime $m_\chi \gg M_\phi$, the annihilation cross section is independent of the DM mass and no information on this parameter can be extracted  by analyzing the relic density only. 

In the low mass region, $m_\chi \ll M_\phi$, the annihilation cross sections can be written as\beq
\langle \sigma v \rangle_{gg}^{0^+} \simeq  \frac{32 c_{gg}^2 g_\phi^2 v^2}{\pi \Lambda^2} \left( \frac{m_\chi}{M_\phi} \right)^4; ~
\langle \sigma v \rangle_{gg}^{0^-} \simeq \frac{128 c_{gg}^2 g_\phi^2}{\pi \Lambda^2}\left( \frac{m_\chi}{M_\phi} \right)^4 ,  ~~  m_\chi \ll M_\phi
\eeq
From these simple expressions, one can see that at low $m_\chi$, $g_\phi$ should be large to allow for a significant annihilation cross section and  avoid the overabundance of DM particles. In turn,  for $m_\chi \approx 300$ GeV in the scalar and $m_\chi \approx 50$ GeV  in the pseudoscalar cases, $g_\phi \gsim 1 $ and we enter  in 
a non-perturbative regime. Moreover, in this range, the DM particles are light enough
for the resonance to decay invisibly, hence increasing its total width. The limit corresponding to a total width of 60 GeV is shown in Fig.~2 by the 
dot-dashed line above which any value of the coupling is excluded as it leads to 
a too large width for the resonance. This is similar to the SM--like 125 GeV  
Higgs portal scenario \cite{Hportal}, which excludes too light DM particles, 
$m_\chi \lsim \frac12 M_h \approx 62$ GeV from its invisible width.

\begin{figure}
\begin{center}
\vspace*{2mm}
\mbox{\hspace*{-4mm} 
 \includegraphics[width=0.48\linewidth]{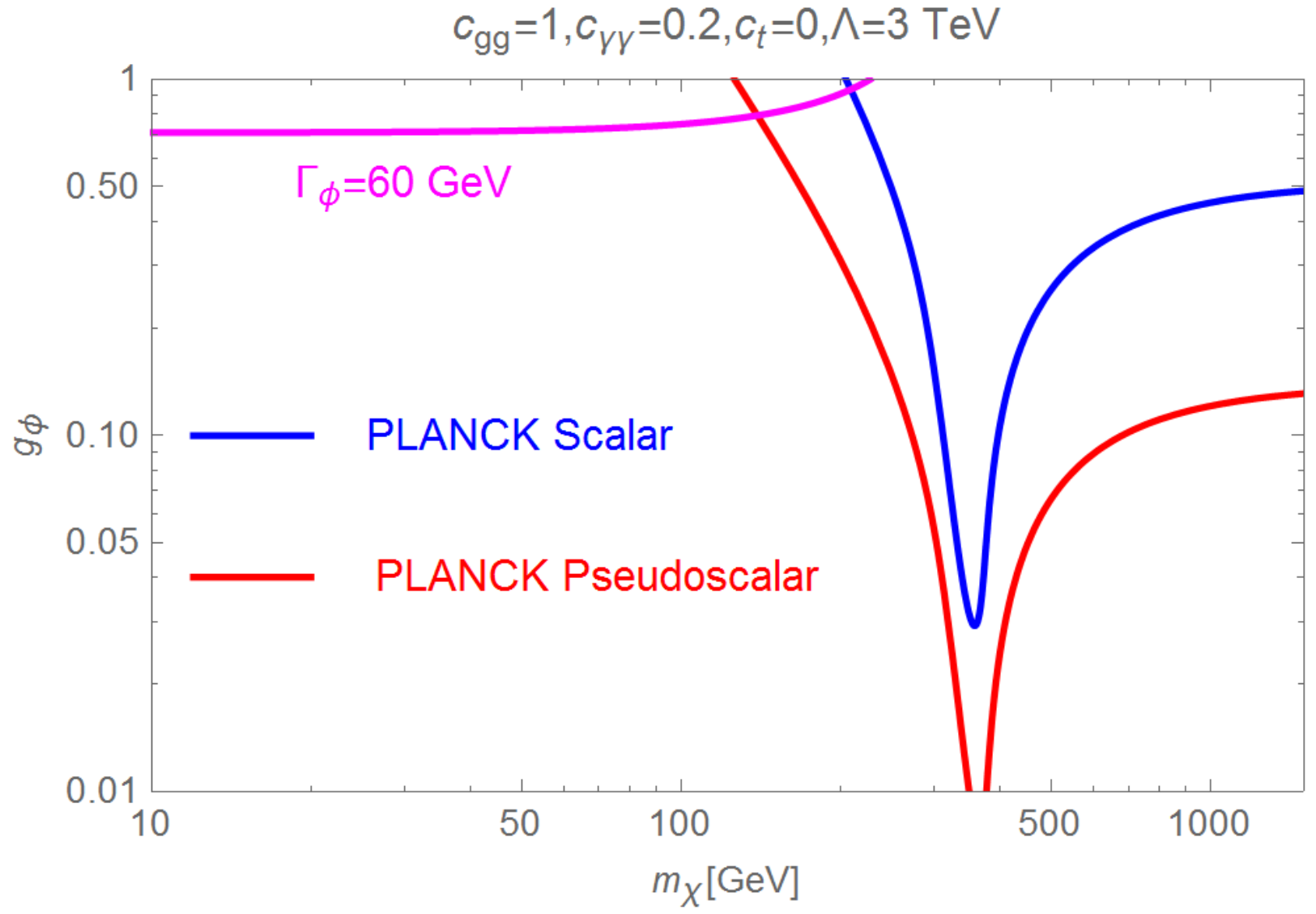}~~
 \includegraphics[width=0.48\linewidth]{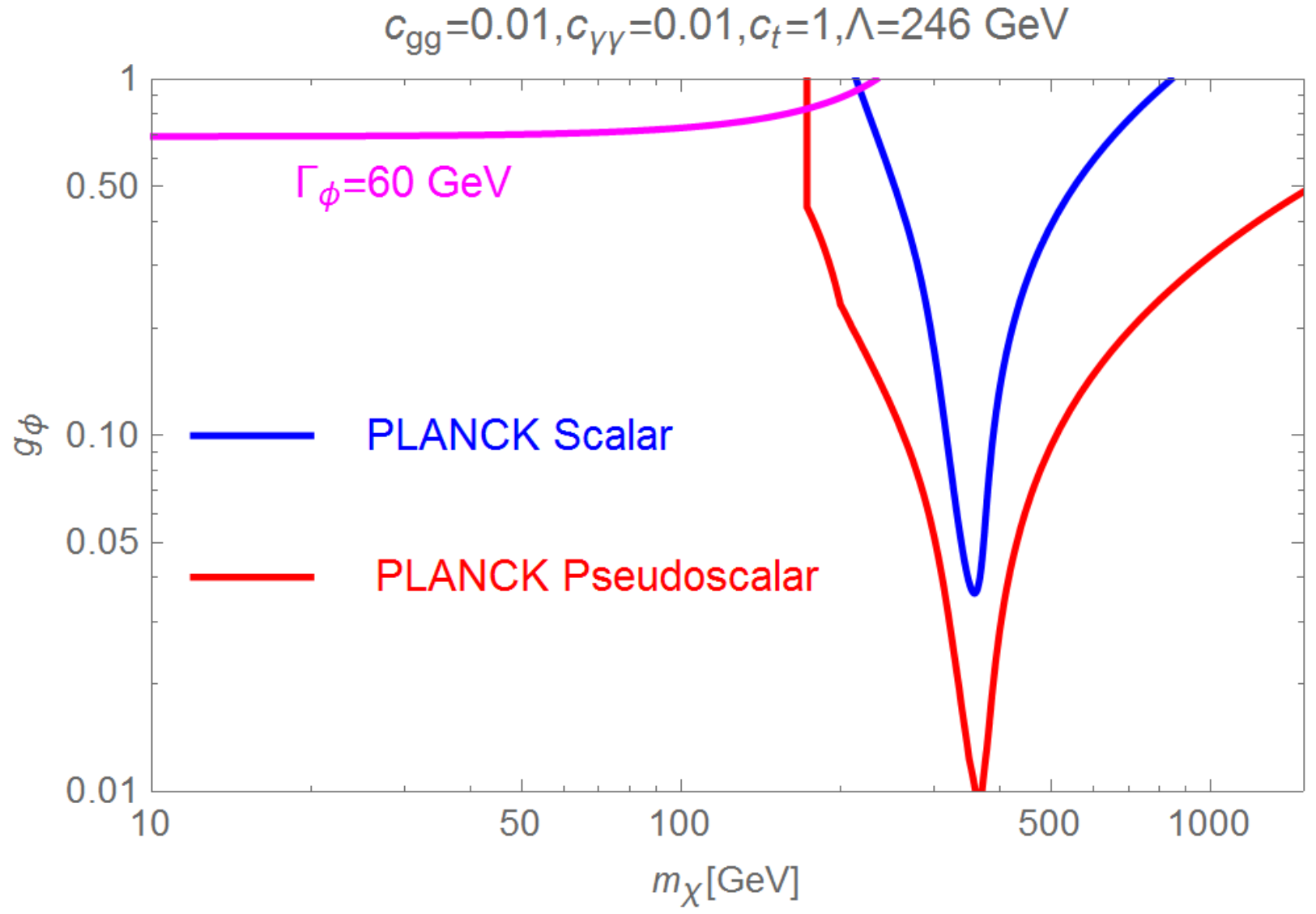}
}
\vspace*{-2mm}
\caption{\small The parameter space in the plane [$m_\chi$, $g_\phi$] 
that is allowed by the relic density measurement for a resonance $M_\phi=750$ 
GeV and $c_i$ and $\Lambda$ reported on top of the figures. In red 
and in blue are shown, respectively, the scalar and pseudoscalar cases. 
We also plot the constraint from a total width of 60 GeV in magenta.}.
\end{center}
\label{Fig:omega}
\vspace*{-11mm}
\end{figure}

The last mass region is the pole region, where $M_\phi \simeq 2 m_\chi$, in which
the annihilation cross section is enhanced and one needs lower values of $g_\phi$ to 
fulfil the WMAP/PLANCK constraint, similarly to the SM Higgs--portal case mentioned above.This region is very interesting as there, the DM mass is fixed to $m_\chi
\approx 350$ GeV and the only free parameter in the DM sector would be the 
coupling $g_\phi$ which is very small,   $g_\phi \approx 10^{-1}$ for scalar exchange 
and $g_\phi \approx 10^{-2}$ for pseudoscalar exchange 

Let us finally discuss the case in which the resonance $\phi$ has couplings to SM fermions. 
If one adds to the Lagrangians ${\cal L}_0$ an effective fermionic coupling,  one generates new DM annihilation channels with two SM fermions. The dominant one should naturally be the channel with top quarks as a result of the large Yukawa coupling. Also this annihilation channel has been consistently included in micrOmegas and then considered in our analysis.

For high values of the scale $\Lambda$, as in the scenario considered in eq.~(\ref{Eq:parameters}), the contribution of DM annihilation into $t \bar t$ pairs is suppressed by a factor $m_t/\Lambda$. Scanning on the allowed region of the parameter space, one finds that the $\chi \chi \to t \bar t$ final state is never dominant and its rates contributes at the level of at most a few percent to the relic density.

The situation is completely different for a low scale, as it is the case in the scenario of  
eq.~(\ref{Eq:parameters2}) in which we have $\Lambda=246\,\mbox{GeV}$, and  the annihilation cross section of the DM particles is dominated by the $\chi \chi \to t \bar t$ channel. In this scenario, as can be seen from the right--hand panel of Fig.~2, small values of the coupling $g_\phi$ are needed in the pseudoscalar case. Indeed, in the $m_\chi < M_\phi$ regime the cross sections in the scalar and pseudoscalar cases can be approximated by
\begin{equation}
\langle \sigma v \rangle_{\bar t t}^{0^+} \approx \frac{g_\phi^2 m_t^2 m_\chi^2 v^2}{4 \pi v_h^2 M_\phi^4} {\left(1-\frac{4 m_t^2}{m_\chi^2}\right)}^{3/2} ,\,\,\,\,
\langle \sigma v \rangle_{\bar t t}^{0^-} \approx \frac{g_\phi^2 m_t^2 s}{4 \pi v_h^2 M_\phi^4} \sqrt{1-\frac{4 m_t^2}{m_\chi^2}} 
\end{equation}
where we have retained the dependence on the top quark mass as it is of order 
of $m_\chi$.  Instead, for $m_\chi > M_\phi$, one simply has 
\begin{equation}
\langle \sigma v \rangle_{\bar t t}^{+} \approx \frac{g_\phi^2 m_t^2 v^2}{64 \pi v_h^2 m_\chi^2}, \,\,\,\, \langle \sigma v \rangle_{\bar t t}^{-} \approx \frac{g_\phi^2 m_t^2}{16 \pi v_h^2 m_\chi^2}
\end{equation}
Similarly to the case in which the annihilation channel into gluons was dominating, 
we have a velocity suppressed cross section in the case of a scalar $\phi$ and an s--wave dominated cross section for the pseudoscalar. Contrary to the former scenario, the annihilation cross section retains a dependence on the DM mass even in the $m_\chi 
> M_\phi$ regime, though.

%\subsubsection*{3.2 Direct detection}

Let us now discuss direct detection. In the case of a scalar mediator, direct detection is accounted for by spin independent interactions which are described, in the non relativistic limit,  by the following Lagrangian~\cite{DelNobile:2013sia}:
\begin{eqnarray}
{\mathcal{L}}_{\chi N}=g_\phi \phi \bar \chi \chi+ g_{\phi N N} \phi \bar N N 
\hspace*{5cm} \\ 
\label{eq:cnn}
g_{\phi N N}=\sum_{f=u,d,s}c_f \frac{m_N}{\Lambda} f^N_{Tf}+\frac{2}{27}f^N_{TG}\left(\sum_{f=c,b,t}c_f \frac{m_N}{\Lambda}-\frac{12 \pi}{\alpha_s}c_{gg}\frac{m_N}{\Lambda}\right)
\end{eqnarray} 
where the factor $12 \pi/\alpha_s$ has been inserted in order to compensate the difference in the coefficient of $\phi G_{\mu \nu}^a G^{\mu \nu,\,a}$ given in~Ref.~\cite{DelNobile:2013sia}. The spin--independent  cross section is then given by
\begin{equation}
\sigma_{\chi p}^{\rm SI}=\frac{\mu_\chi^2}{\pi}\frac{g_{\rm \phi}^2}{M_\phi^2}g_{\phi NN}^2
\end{equation}
In the case $c_f=0$ for all the quarks, we have
\begin{equation}
\sigma_{\chi p}^{\rm SI} \approx 10^{-43}\,{\mbox{cm}}^2 {\left(\frac{M_\phi}{750\,\mbox{GeV}}\right)}^{-4} {\left(\frac{\Lambda}{1\,\mbox{TeV}}\right)}^{-2}
\end{equation} 

\begin{figure}[!h]
\hspace*{-2mm} 
\begin{center}
\mbox{\hspace*{-5mm} 
 \includegraphics[width=0.48\linewidth]{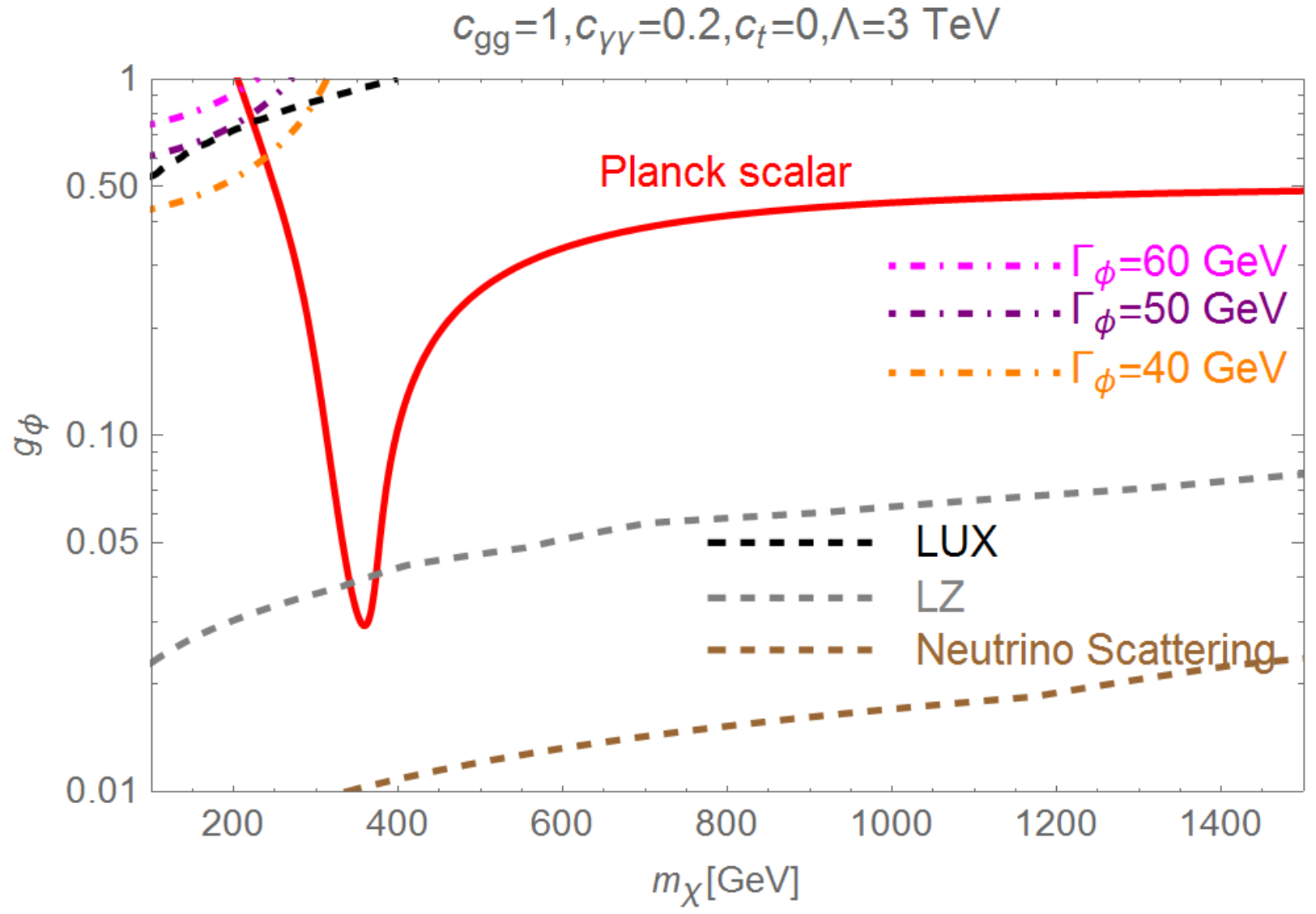}~~
 \includegraphics[width=0.48\linewidth]{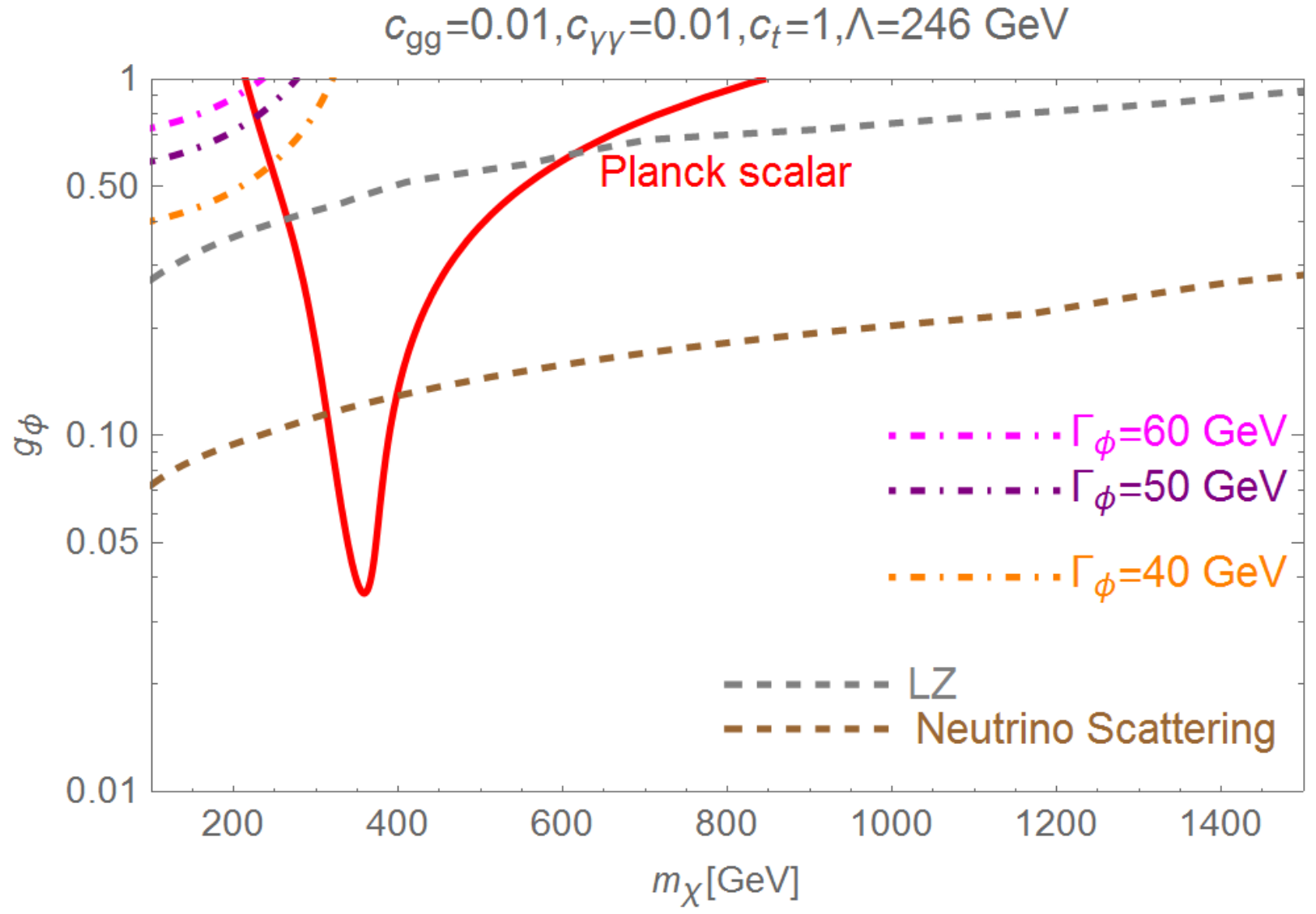}
}
\hspace*{-2mm} 
\caption{\small
Direct detection prospects in the case of a scalar $\phi$ for two sets of parameters  reported on top of the figures, including the LZ prospects \cite{LZ}. The red lines are for measured DM relic density and the isocontours for a a given total width 
of the resonance are shown.}
\end{center}
\label{Fig:detection_gg}
\vspace*{-3mm}
\end{figure}

The limits/prospects from direct detection are compared with the requirement of the correct relic density in Fig.~3 for two sets of the $c_{\rm gg},c_f,\Lambda$ parameters. In the left panel we have considered $c_{\rm gg}=1$, $\Lambda=3\,\mbox{TeV}$ and $c_f=0$ for all  quarks, while in the right panel we have taken $c_{\rm gg}=0.01$, $c_t=1$ (the coupling with other SM fermions have been kept fixed to zero) and $\Lambda=246\,\mbox{GeV}$. In the first case, we notice a moderate impact from the current limit set by LUX, which excludes the region $m_\chi \lesssim 200\,\mbox{GeV}$ and a width of the scalar state greater than 50 GeV. On the contrary, the future multi tons experiment LZ \cite{LZ}, which is supposed to begin commissioning in 2017, will fully probe this configuration of the parameters. In the second scenario, the a cancellation between the contributions relative to the couplings $c_t$ and $c_{\rm gg}$, eq.~(\ref{eq:cnn}), render the detection prospects more modest. Indeed only LZ is capable to partially probe this configuration and the resonance region, namely $m_\chi \sim M_\phi/2$ is even below the neutrino scattering limit~\cite{Billard:2013qya}, where  the sensitivity will reach the level of detection of atmospheric neutrino.  

The scattering cross section of Majorana fermions exchanging a pseudoscalar particle, being proportional to the velocity (which is about 200 km/s around the earth), gives instead null detection prospects, see Ref.~\cite{coy} for instance. 

%\subsubsection*{3.3  Indirect detection}

Finally, let us make a few comments on indirect detection.  The detection of the DM particle  through the observation of its annihilation in the Galactic Center  or in nearby dwarf galaxies is one of the most promising ways. Nevertheless, the latest  analyses of FERMI \cite{Ackermann:2013yva} and HESS \cite{Hess} do not exhibit any signal so far. The only testable case is obviously when  the mediator particle is a pseudoscalar which, in contrast to the case where the annihilation cross section goes through the  exchange of a scalar, is not velocity suppressed; see eq.~(\ref{Eq:sigv}).  We show in Fig.~3 the limits obtained by 
the FERMI experiment from the latest observation of dwarf galaxies \cite{Ackermann:2015zua}.
Again we have used our two usual scenarios for illustration and displayed the 
regions favored by the measurement of the relic density and a resonance width of 60 GeV.  
From the figure, one sees that the interesting region should be soon probed by the collaboration, within the next few years,  just by accumulating statistics.

\begin{figure}[!h]
\begin{center}
\mbox{\hspace*{-5mm} 
\includegraphics[width=0.46\linewidth]{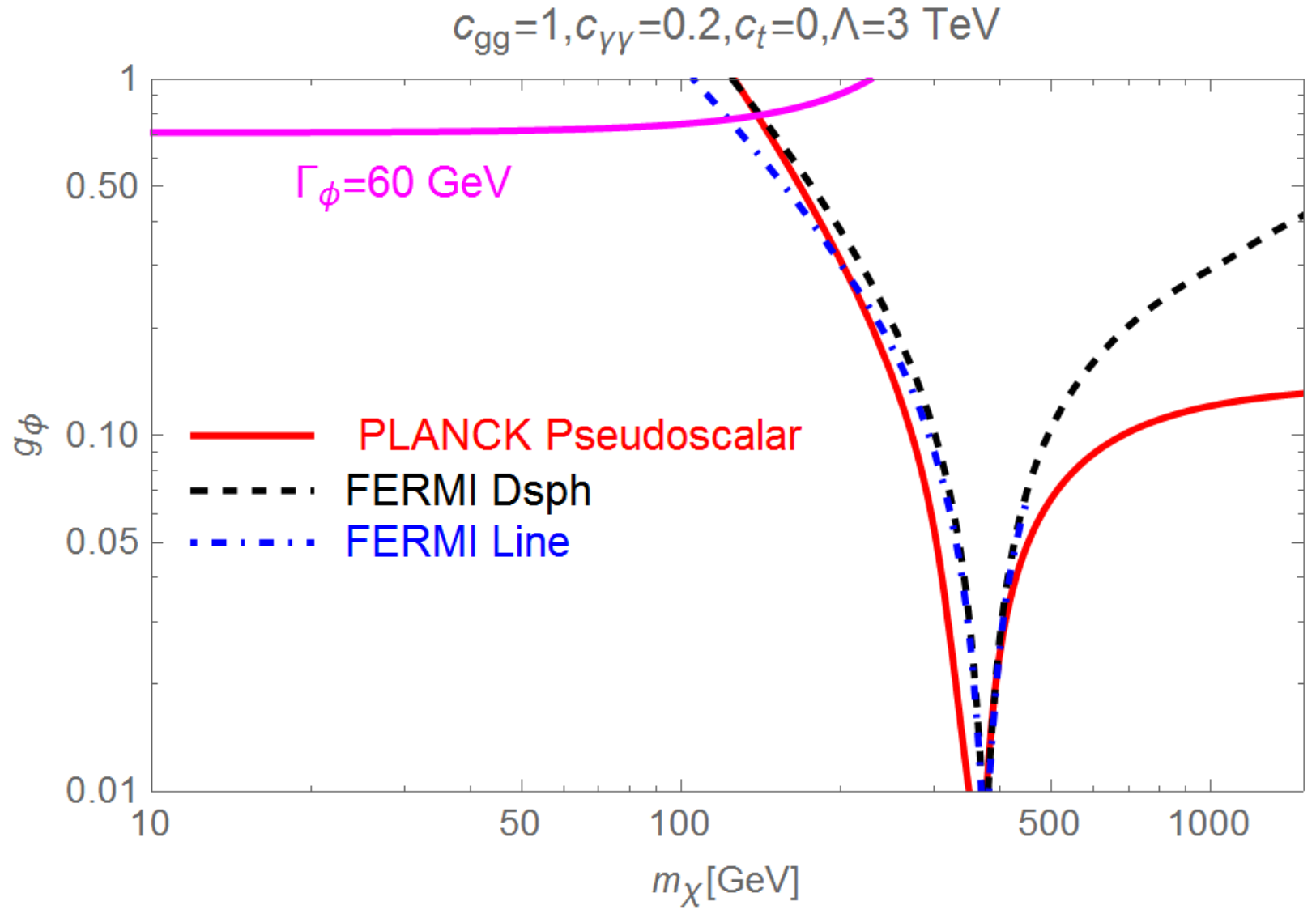}
\includegraphics[width=0.46\linewidth]{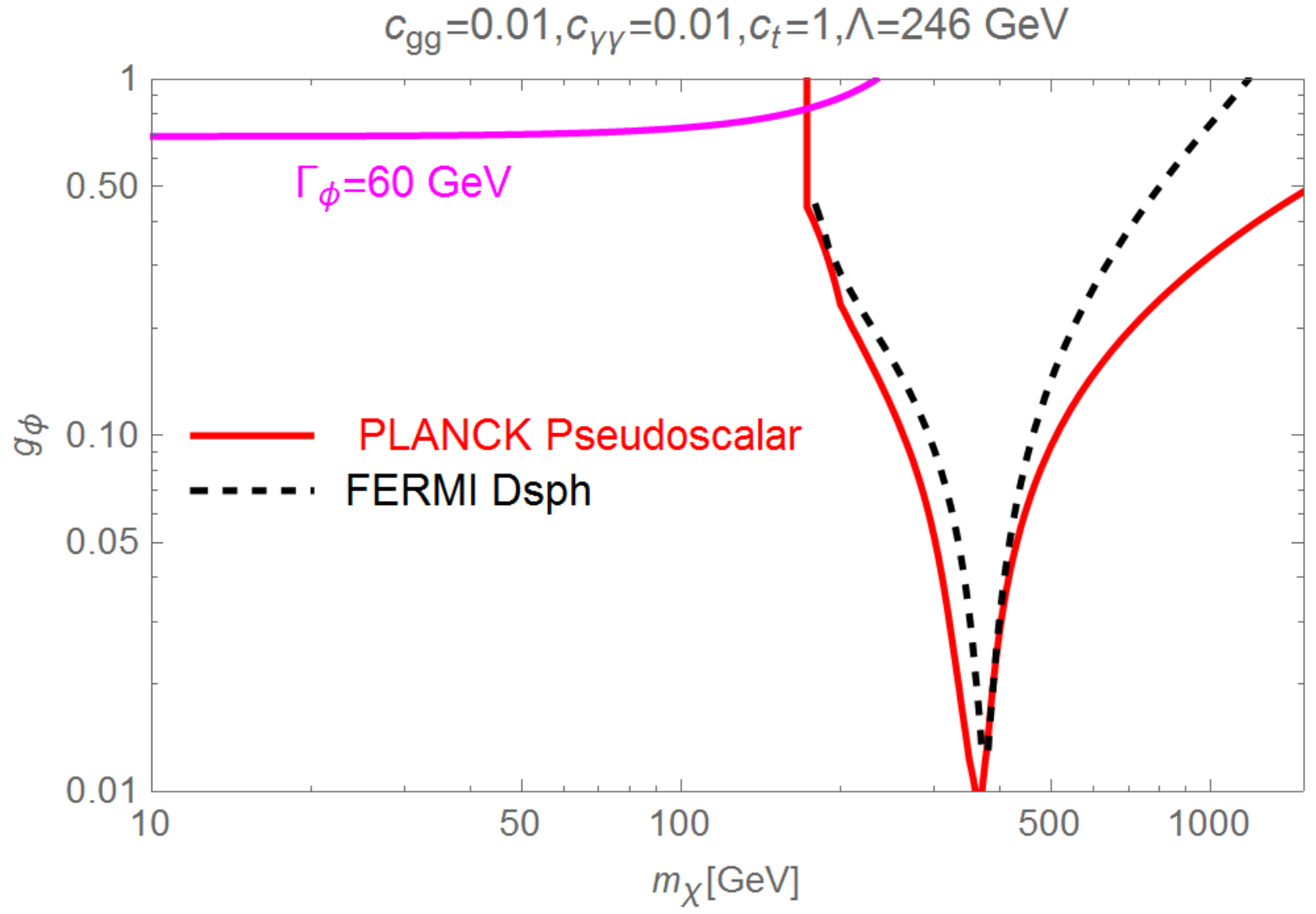}
}
\vspace*{-2mm}
\caption{\small
Indirect detection prospects in the case of a pseudoscalar $\phi$ resonance in the 
$[m_\chi, g_\phi]$ plane for two sets of parameters  reported on top of the figures. 
The red lines are for measured DM relic density and the isocontours for a given total width 
for the resonance are shown.
}
\end{center}
\label{Fig:detection}
\vspace*{-3mm}
\end{figure}

One characteristic of the model would be the presence of a monochromatic photon
generated by $\langle \sigma v \rangle_{\gamma \gamma}$ above a continuous spectrum generated
by $\langle \sigma v \rangle_{ZZ, gg, ..}$; it is reminiscent to effective constructions studied in Ref.~\cite{Chu:2012qy} for instance. The ratio of the monochromatic signal over the background is indeed determined completely once the $c_i$ parameters
of the resonance are fully identified. In the scenario of a high scale $\Lambda$, as described for example by the scenario of  eq.~(\ref{Eq:parameters}), the annihilation cross section of the DM is dominated by gauge boson final states and we have roughly $\langle \sigma v \rangle_{\gamma \gamma} = c_{\gamma \gamma}^2 / (8 ~c_{gg}^2) \times \langle \sigma v \rangle_{gg}$. We have reported in the left panel of Fig.~5 the limit from the most recent searches of Galactic gamma-ray lines~\cite{Ackermann:2015lka}. As evident,  this last limit gives analogous constraints with respect to dwarf galaxies. In the case described by  
eq.~(\ref{Eq:parameters2}) with a low scale $\Lambda$, the branching fraction of the two--photon final state is negligible and we obtain relevant limits only from the gamma--ray continuum.

\subsection*{4. Conclusion}

In this note, we performed a quick analysis of the implications for dark mater of the 
excess that was observed by the ATLAS and to a lesser extent the CMS collaborations in the diphoton invariant mass spectrum and which points to the existence of a resonance  of mass of
about 750 GeV. Interpreting the resonance as the scalar or pseudoscalar mediator of the annihilation of DM into standard particles, we have shown that the LHC data are perfectly  compatible with the WMAP and PLANCK satellite measurements. The correct relic abundance
can be obtained  for reasonable masses of the DM particle, $m_\chi \approx 0.1$--1 TeV,  
and couplings to the resonance, $g_\phi \approx 10^{-1}$. 

We have then  discussed the prospects for detecting the DM particles in astrophysical experiments. The  prospects  for indirect detection are reasonably optimistic in the case of a pseudoscalar resonance. In turn, one should be quite optimistic for the direct detection mode in the case of a scalar mediator, when couplings with the top are neglected, as the cross section is within the reach of LZ. On the contrary, in presence of couplings of the scalar with both gluons and top quarks and for a low energy scale $\Lambda$, direct detection prospects appear less encouraging. In any case, the DM models incorporating the diphoton resonance observed at CERN, if indeed real,  are testable in the generation of detectors that are presently being  built.\bigskip

{\bf Acknowledgements:} Y.M. would like to thank Emilian Dudas for fruitful discussions and Sasha Pukhov for resolving technical issues in micrOmegas. A.D. thanks Giovanni Lamanna for
discussions. The authors also warmly thank Paolo Panci,  Diego Redigolo and Andreas Goudelis for useful comments and a fruitful correspondence.
This work is supported by the ERC advanced grant Higgs@LHC. Y.M.  is supported by the European Union FP7 ITN INVISIBLES and MassTeV, the Spanish MICINN's Consolider-Ingenio 2010
Programme under grant Multi-Dark {CSD2009-00064}, the contract {FPA2010-17747}, the France-US PICS no. 06482 and the LIA-TCAP of CNRS.  G.A. acknowledges the University of Genova for the warm hospitality during part of the completion of this project and A.D. the CERN theory unit for hospitality.

\end{document}